\documentclass[onecolumn,aps,nofootinbib,a4paper]{revtex4} 

\usepackage[english]{babel}
\usepackage[utf8]{inputenc}

\usepackage{amsmath,amssymb}
\usepackage{graphicx}

\usepackage{kpfonts}
\newenvironment{veryoldstyle}{%
  \normalfont
}{}

\usepackage{color}

\usepackage{hyperref}

\begin{document}

\title{Vacuum Friction}

\author{Stephen M. Barnett}
\affiliation{School of Physics and Astronomy, University of Glasgow, Glasgow G12~8QQ, United Kingdom}

\author{Matthias Sonnleitner}
\affiliation{School of Physics and Astronomy, University of Glasgow, Glasgow G12~8QQ, United Kingdom}

\begin{abstract}
We know that in empty space there is no preferred state of rest.  This is true both in special relativity but also in 
Newtonian mechanics with its associated Galilean relativity.  It comes as something of a surprise, therefore, to 
discover the existence a friction force associated with spontaneous emission.  The resolution of this paradox
relies on a central idea from special relativity even though our derivation of it is non-relativistic.  We examine the
possibility that the physics underlying this effect might be explored in an ion trap, via the observation of a superposition
of different mass states.
\end{abstract}

\maketitle

{\it ``The term `holistic' refers to my conviction that what we are concerned with here is the fundamental interconnectedness of all things.  I do not concern myself with such petty things as fingerprint powder, telltale pieces of pocket fluff and inane footprints.
I see the solution to each problem as being detectable in the pattern and web of the whole.  The connections between causes
and effects are often much more subtle and complex than we with our rough and ready understanding of the physical world
might naturally suppose."} 

\hfill Douglas Adams \cite{Adams}

\section{Introduction}

It has long been appreciated that the optical Doppler shift could be used to cool a gas of atoms \cite{HaenschSchawlow} or a
trapped ion \cite{WinelandDehmelt,Mavadia}.  The essential idea is that a narrow-line laser tuned below a resonant transition
frequency for the atom will be absorbed, preferentially if the atom is moving towards the light source because of the Doppler shift 
and hence be slowed down so as to accommodate the momentum of the photon \cite{AdamsRiis}.  If the absorption of a laser 
photon is followed by spontaneous emission a further photon can be absorbed and, after a number of cycles, the average 
velocity of the atom is reduced.  If the single laser beam is supplemented by five more then Doppler cooling can be achieved
in three dimensions \cite{Chu} and an atom feels an average frictional force ${\bf F} = -\alpha{\bf v}$, where ${\bf v}$ is the 
velocity of the atom.  These ideas were the seed from which the field of laser cooling and trapping, and much else besides, 
has flowered \cite{Ashkin}.

\begin{figure}
   \centering
   \includegraphics[width=0.7\textwidth]{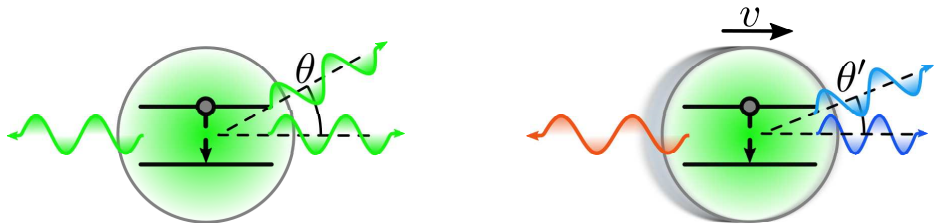}
   \caption{\label{fig:_emission}%
   Illustration of the (spontaneous) emission process of an atom at rest (left figure) or moving at speed~$v$ (right). An atom at rest emits photons of the frequency $\omega_0$ in all directions~$\theta$ so that the average recoil is zero. A moving atom emits photons of frequency $\omega_0'\approx \omega_0 \big(1 + \tfrac{v}{c} \cos \theta' \big)$ in direction $\cos\theta' \approx \cos\theta + \tfrac{v}{c}\big(1-\cos^2\theta\big)$. Integrating the recoil over all directions results in a non-zero change of the atom's momentum.}
\end{figure}

Hidden within the combination of the optical Doppler shift and the interaction between light and atoms is a paradox, which we
have identified recently \cite{VacFric}.  The point is simply made: an excited atom in an otherwise empty region of space can
return to its ground state by the spontaneous emission of a photon.  In doing so it receives a recoil so as to conserve momentum;
if the emitted photon has momentum $\hbar{\bf k}$ then the momentum of the atom changes, correspondingly, by 
$-\hbar{\bf k}$.  If the atom is stationary then the essentially isotropic nature of spontaneous emission means that there the
net or \emph{average} change in the momentum is zero.  If the atom is moving, however, then a photon emitted in the direction
of motion of the atom will have, by virtue of the Doppler shift, a higher frequency and hence a higher momentum than one
emitted in the opposite direction, as depicted in Fig. \ref{fig:_emission}.  If we take the average over these events we are led to 
a net reduction in the momentum of the atom in the emission process.  This reduction, moreover, is proportional to the velocity 
of the atom.  In short, we have a friction force associated with the spontaneous emission event.  Yet the existence of a force in 
one frame that does not exist in another seems to be at odds with both the Galilean and Einsteinian principles of relativity.  
Hence we have a paradox.

Our earlier publication on this problem was necessarily somewhat formal and rigorous \cite{VacFric}.  Our aim here is to present 
the key ideas in a more physical fashion.  We shall find that the resolution of the paradox lies in a key idea from special 
relativity, but the remarkable feature of our analysis is that it requires ideas familiar only from \emph{non-relativistic} physics.  We 
consider also the possibility that the central idea involved in this resolution might be observable in an ion-trap experiment.

\section{Vacuum friction: a physical derivation}

Let us consider the spontaneous emission by an atom via an electric-dipole transition.  A simple calculation based on
Fermi's golden rule gives the spontaneous emission rate \cite{Rodney}
\begin{equation}
\Gamma = \frac{\omega_0^3|{\bf d}|^2}{3\pi\epsilon_0\hbar c^3} ,
\end{equation}
where ${\bf d}$ is the transition dipole matrix element and $\omega_0$ is the atomic transition frequency.  If the atom is
moving then there will be a change to this decay rate but only that required by the time-dilation of special relativity,
as may be confirmed by direct calculation \cite{Jim}.  This effect is of second order in the velocity and will not be of 
concern to us in this paper; we are interested in those effects that arise at first order in $v/c$.

We can apply physical reasoning to arrive at least at an approximate form for the paradoxical friction force and it is 
instructive to follow this approach.  To this end we introduce the idea of a spontaneous emission rate into an infinitesimal 
solid angle, 
\begin{equation}
d\Gamma = \Gamma \frac{d\Omega}{4\pi}  , 
\end{equation}
which is simply the spontaneous emission rate multiplied by the probability of emission into the solid angle $d\Omega$.
If we integrate this over all directions of emission then we recover the full spontaneous emission rate, 
$\int d\Gamma = \Gamma$.

If the atom is stationary then the emitted photon will carry away an amount of momentum given by 
$\hbar k_0 = \hbar\omega_0/c$.  The net or average recoil force on our atom is simply
\begin{eqnarray}
{\bf F} &=& -\int \hbar {\bf k}_0 d\Gamma P_e(t) \nonumber \\
&=& -\int \hbar {\bf k}_0 d\Gamma e^{-\Gamma t}  \nonumber \\
&=& 0 ,
\end{eqnarray}
where $P_e(t)$ is the probability that the atom is still in the excited state at time $t$.  The zero value of this net force
is a direct consequence of the lack of a preferred direction for the spontaneous emission which is, itself, a consequence
of the isotropy of empty space.

If the atom is moving then this motion selects for us a direction in space.  Let us take this direction to define the $z$-axis
so that the atomic velocity is ${\bf v} = v\hat{\bf z}$.  Because of this motion the Doppler effect means that the frequency of the 
emitted photon will depend on the direction in which it is emitted.  If the angle between the direction of the emitted photon
and the $z$ axis is $\theta$ then the observed frequency of the photon will be 
\begin{equation}
\omega_0' = \omega_0\left(1 + \frac{v}{c}\cos\theta\right) .
\end{equation}
If we insert this Doppler-shifted frequency into our expression for the net force we find a first tentative expression for the
force due to spontaneous emission on a moving atom:
\begin{eqnarray}
{\bf F}_1 &=& -e^{-\Gamma t}\int \hbar\frac{\omega_0}{c}\left(1 + \frac{v}{c}\cos\theta\right) 
\cos\theta d\Gamma  \, \hat{\bf z} \nonumber \\
&=& -\frac{1}{3}e^{-\Gamma t}\frac{\hbar\omega_0}{c^2}\Gamma \frac{v}{c} \, \hat{\bf z}  \nonumber \\
&=& -\frac{1}{3}e^{-\Gamma t}\frac{\hbar\omega_0}{c^2}\Gamma \, {\bf v} ,
\end{eqnarray}
which has the characteristic form of a friction force.  Here we have exploited the rotational symmetry about the $z$-axis
to infer directly that the force must be parallel to this axis.\footnote{We could include, explicitly, the dipole radiation pattern 
but, for simplicity, consider a spatially averaged dipole.}

Is there anything we have left out and, in particular, is there any non-relativistic effect that is missing?  The answer,
of course, is yes and the additional feature is the aberration of light due to motion.  This is a familiar element of 
special relativity where the constancy of the speed of light leads to a modification 
of the angles measured in different frames and this requires us to replace the cosines and sines by
\cite{Rindler}
\begin{eqnarray}
\label{costransform}
\cos\theta \rightarrow \cos\theta' &=& \frac{\cos\theta + v/c}{1 + (v/c)\cos\theta} \nonumber \\
\sin\theta \rightarrow \sin\theta' &=& \frac{\sin\theta}{\gamma(1 + (v/c)\cos\theta)} ,
\end{eqnarray}
where $\gamma$ is the usual Lorentz factor, $\gamma = (1-v^2/c^2)^{-1/2}$.
It is important to note, however, that this idea is much older than relativity.  Indeed it was first noted by Bradley in 1729 
\cite{Bradley} in the response to the appearance of consistent discrepancies in the measurement of stellar parallaxes.
Bradley's explanation for this was a modification in the perceived angles due to the finite value of the speed of light:

\vskip0.2cm

\begin{veryoldstyle}
\noindent ``And in all Cases=, the Sine of the Difference between the real and visible Place of the Object, will be to the 
Sine of the visible  Inclination of the Object to the Line in which the Eye is= moving, as= the Velocity of the Eye to the Velocity of Light." \end{veryoldstyle}
\cite{Bradley}.

\vskip0.2cm

\noindent Putting these words into mathematical form we arrive at 
\begin{equation}
\sin(\theta - \theta') = \frac{v}{c}\sin\theta' ,
\end{equation}
which is readily recovered from the relativistic expression (\ref{costransform}) in the limit of small velocity.  Rather than work with
Bradley's expression it is more transparent (although strictly equivalent) to work with the relativistic formula
and restrict ourselves to low velocities by working to first order in $v/c$.  The aberration means that the angle
between the $z$-axis and the direction of emission of the photon, the term $\cos\theta$ in our expression
for the force should be replaced by $\cos\theta'$.  If we make this substitution then we arrive at the correct
expression for the net force:
\begin{eqnarray}
\label{Force}
{\bf F} &=& -e^{-\Gamma t}\int \hbar\frac{\omega_0}{c}\left(1 + \frac{v}{c}\cos\theta\right) \cos\theta' d\Gamma 
\, \hat{\bf z} \nonumber \\
&=& -e^{-\Gamma t}\int \hbar\frac{\omega_0}{c}\left(\cos\theta + \frac{v}{c}\right) d\Gamma 
\, \hat{\bf z} \nonumber \\
&=& -e^{-\Gamma t}\frac{\hbar\omega_0}{c^2}\Gamma {\bf v} .
\end{eqnarray}
As a check of this idea we can work with the primed angle, $\theta'$, and note that this involves a change in the
integration variable
\begin{eqnarray}
d(\cos\theta') &=& \frac{d(\cos\theta)}{(1 + (v/c)\cos\theta)^2} \nonumber \\
\Rightarrow
d\Omega &=& d\Omega' (1 + (v/c)\cos\theta)^2 .
\end{eqnarray}
Evaluating the net force in this manner and again working to first order in $v/c$ gives
\begin{eqnarray}
{\bf F} &=& -e^{-\Gamma t}\int \hbar\frac{\omega_0}{c}\left(1 + \frac{v}{c}\cos\theta\right) \cos\theta' d\Gamma' 
 \left(1 + \frac{v}{c}\cos\theta\right)^2\, \hat{\bf z} \nonumber \\
&=& -e^{-\Gamma t}\int \hbar\frac{\omega_0}{c}\left(1 + 3\frac{v}{c}\cos\theta'\right) \cos\theta' d\Gamma' 
\, \hat{\bf z} \nonumber \\
&=& -e^{-\Gamma t}\frac{\hbar\omega_0}{c^2}\Gamma {\bf v} ,
\end{eqnarray}
where we have again worked to first order in $v/c$.  We have arrived at the same expression by using two 
sets of angles, $\theta$ and $\theta'$.  In the first we take account of the aberration by transforming the angle
of emission, but in the second it is the solid angle into which the emission occurs that is transformed.

The above analysis considered isotropic emission and one might wonder if our simple result changes for a more general emission pattern. A general emission pattern is defined by a quantity $\gamma(\mathbf{k})$, the rate at which light with wavevector 
$\mathbf{k}$ is emitted, if the atom is in the excited state. The total decay rate is thus
\begin{equation}\label{decayrate}
	\Gamma = \int \gamma(\mathbf{k}) d^3\mathbf{k} ,
\end{equation}
where we integrate over all directions and frequencies, $d^3\mathbf{k} = \omega^2 d\omega d\Omega$. As before we express the wavevector in spherical coordinates, $\mathbf{k}= \omega/c \left( \sin\theta \cos\phi, \sin\theta \sin\phi, \cos\theta \right)^T$. In the rest frame of the emitter we get a recoil force
\begin{equation}\label{force_restframe}
	\mathbf{F} = -e^{-\Gamma t} \int d^3\mathbf{k} \hbar \mathbf{k} \gamma(\mathbf{k}) .
\end{equation}
For a spontaneously decaying atom $\gamma(-\mathbf{k})=\gamma(\mathbf{k})$ and this net recoil force in the rest frame is, of course, zero.

In a frame where the emitter is moving at a velocity ${\bf v}=v \hat{\bf z}$, we express $\mathbf{k}'$ using the Doppler shift and aberration to first order,
\begin{eqnarray}
	\mathbf{k}' &\simeq& \frac{\omega}{c}\left(1 + \frac{v}{c}\cos\theta\right)
		\begin{pmatrix}
			\sin\theta' \cos\phi \\
			\sin\theta' \sin\phi \\
			\cos\theta'
		\end{pmatrix} ,
		\nonumber \\
		&=& \mathbf{k} + \frac{\omega\mathbf{v}}{c^2} ,
		\label{ktransform}
\end{eqnarray}
where we used the aberration formula, equation~(\ref{costransform}), to arrive at the last line.  We could also have used Bardley's non-relativistic expression as we are working only to first order in $v/c$.

It is not necessary to give an explicit transformation for the general emission pattern $\gamma(\mathbf{k})$, as we know that the amount of radiation emitted into each volume element must be invariant, so that
\begin{equation}
	\gamma(\mathbf{k}) d^3\mathbf{k} = \gamma'(\mathbf{k}') d^3\mathbf{k}' . 
\end{equation}
As the decay rate $\Gamma$ only changes with second order in velocity, we can express the force in the moving frame in terms of the unprimed quantities,
\begin{eqnarray}
	\mathbf{F} &=& -e^{-\Gamma t} \int d^3\mathbf{k}' \hbar \mathbf{k}' \gamma'(\mathbf{k}') ,
		\nonumber\\
		&=& -e^{-\Gamma t} \hbar \int d^3\mathbf{k} \gamma(\mathbf{k}) \left( \mathbf{k} + \omega \mathbf{v}/c^2\right) ,
		\nonumber\\
		&=&  - e^{-\Gamma t} \frac{\hbar \mathbf{v}}{c^2} \int d^3\mathbf{k} \gamma(\mathbf{k}) \omega 
		\label{force_movingframe_general}
\end{eqnarray}
 and we again find a friction force proportional to the velocity of the atom.

The spontaneous decay rate depends on the density of modes at the transition frequency $\omega_0$ and so we write $\gamma(\mathbf{k}) = \delta(\omega - \omega_0) \widetilde{\gamma}(\Omega)$ where we can set $\int d\Omega \widetilde{\gamma}(\Omega) = \Gamma/\omega_0^2$ such that (\ref{decayrate}) is satisfied. It follows that the average or net force is
\begin{eqnarray}
	\mathbf{F} &=& - e^{-\Gamma t} \frac{\hbar \mathbf{v}}{c^2} \int d\omega \omega^3 \delta(\omega - \omega_0) \int d\Omega \widetilde{\gamma}(\Omega) 
	\nonumber \\
		&=&  - e^{-\Gamma t} \frac{\hbar \omega_0}{c^2} \Gamma \mathbf{v}
\end{eqnarray}
in agreement with our earlier expression (\ref{Force}).

This simple expression for the net force ${\bf F}$, which can be obtained more rigorously \cite{VacFric}, has confirmed 
our initial instincts that there is indeed a vacuum friction force, however counterintuitive this conclusion may be.  Our next 
task is to resolve this paradox.

\section{Resolution of the paradox}

If you ask a class of physics students to state Newton's second law of motion, it is likely that the answer you will
get (apart from those smart enough to spot a trap) will be ${\bf F} = m{\bf a}$.  If we apply this to the force we have just
derived then we find
\begin{equation}
\dot{\bf v} = -e^{-\Gamma t}\frac{\hbar\omega_0}{mc^2}\Gamma {\bf v} ,
\end{equation}
where $m$ is the mass of the atom.  This effect is certainly small: it is proportional to the ratio of the photon energy 
to the rest-mass energy of the atom and this ratio is typically of the order $10^{-10}$.  There is an important point of
principle, however, in that if the deceleration exists, whatever its value, then we have a conflict with relativity, 
both of the Einsteinian and Galilean forms.  To emphasise this point we can integrate this equation to find the net change 
in the velocity of the atom:
\begin{equation}
{\bf v}(\infty) = \exp\left(-\frac{\hbar\omega_0}{mc^2}\right){\bf v}(0) \approx \left(1 - \frac{\hbar\omega_0}{mc^2}\right){\bf v}(0) .
\end{equation}
So, if true, this says that the observed speed of the atom is reduced by a simple factor and that this is the case, moreover,
irrespective of the speed of the atom.  So if we see an excited atom in motion then, on average, following the emission
the speed will be reduced but if we are co-moving with the atom then there is no corresponding average change in 
the speed.  This cannot be true.

The resolution of the paradox comes from a surprising place in that it embodies an intrinsically relativistic notion, indeed
perhaps the most famous idea in relativity - the equivalence of energy and mass or inertia.  In the physics class mentioned
above you might find a student who pauses to spot the catch and says, in answer to the question, ``force equals rate of
change of momentum".\footnote{Newton's statement was: \begin{veryoldstyle}
``Mutationem motus= proportionalem esse vi motrici impressae \& fieri secundum lineam rectam qua vis= illa imprimitur."   \end{veryoldstyle} \cite{Newton}, rendered by Ball as ``The change of momentum [per unit time] is always proportional 
to the moving force impressed, and takes place in the direction in which the force is impressed''~\cite{Ball}.}
, ${\bf F} = \dot{\bf p}$.  
They might recall, for example, the classic problem of the motion of a 
space rocket, that burns fuel and, in the process, reduces its mass \cite{Fowles}.  If this is the resolution of the paradox
then we can only infer that the emission of the photon corresponds to a loss of mass by the atom.  Let us see where this
leads us.  If we allow for the possibility that the emission process changes the mass of the atom then Newton's second
law gives us:
\begin{equation}
\dot{m}{\bf v} + m\dot{\bf v} = -e^{-\Gamma t}\frac{\hbar\omega_0}{c^2}\Gamma {\bf v} .
\end{equation}
A change in the average velocity is, as we have seen, paradoxical and suggests that we should set $\dot{\bf v} = 0$.  If we do
this we are left with a simple equation for the rate of change of the mass of the atom:
\begin{equation}
\dot{m} = -e^{-\Gamma t}\frac{\hbar\omega_0}{c^2}\Gamma .
\end{equation}
Which leads directly to the suggestion that the atom is lighter after emitting the photon than when it was prepared initially
in its excited state:
\begin{equation}
m(\infty) - m(0) = -\frac{\hbar\omega_0}{c^2} .
\end{equation}
From the viewpoint of those schooled in special relativity this makes perfect sense: the transition has lowered the total 
energy of the atom by $\hbar\omega_0$ and using the most famous equation in physics, $E = mc^2$, we are led to 
conclude that the ground state atom is indeed lighter than the excited state atom by precisely $\hbar\omega_0/c^2$.  Indeed
very early in the development of relativity, Einstein established this relationship between a change in internal energy and
a change in inertia \cite{Einstein}.  From the viewpoint of special relativity this is entirely unsurprising, of course, given
the close relationship between momentum and energy, which combine as a four-vector.

Taking this view, however, misses our point.  We have employed an entirely 
non-relativistic analysis to arrive at a paradox the only resolution of which seems to imply the necessity of 
a central feature of special relativity.  Perhaps Adams's eponymous hero Dirk Gently was indeed correct and that there is a
fundamental interconnectedness in the physical world and the need for a holistic approach.

\section{Possibility of an ion-trap experiment}

Is there an experiment to be done in order to test these ideas?  As stated, the lack of a change in the average velocity of
a radiating atom would be a difficult, and perhaps not very satisfying, result to establish.  It is indicative of the scale of the
challenge that we would be seeking to find no net change as opposed to a change of perhaps one part in $10^{10}$.  There
is an interesting experimental challenge, however, the demonstration of which would verify the resolution presented above.
This is to show that an atom prepared in an electronically excited state has a higher mass than one in the ground state
and that this difference is given by the transition energy divided by the square of the speed of light.

To show that an excited state is more massive than a ground state requires a mass measurement with a precision of
$\hbar\omega_0/mc^2$ which, for visible light and an atomic mass will be of the order of one part in $10^{10}$.  It is
reasonable to start with the observation that such a measurement, if it is possible, will require a significant period of
time to perform and this means, necessarily that we need to identify an atom with a very long-lived excited state.  We 
note that ion traps have been used to measure atomic masses with a precision of $1-10 keV$ \cite{Blaum}.  Although 
this is roughly three orders of magnitude larger than the typical optical transition energy we are seeking to measure it
is at least encouraging.  Excited ionic states with very long lifetimes are known, moreover, and some of these have been
considered as candidates for frequency standards.  For these reasons it is natural to consider an ion trap
experiment as the way to test this idea.  We shall not propose a specific implementation here but rather present a 
``back of the envelope" assessment of what might be possible with currently available technology.

\begin{figure}
   \centering
   \includegraphics[width=0.5\textwidth]{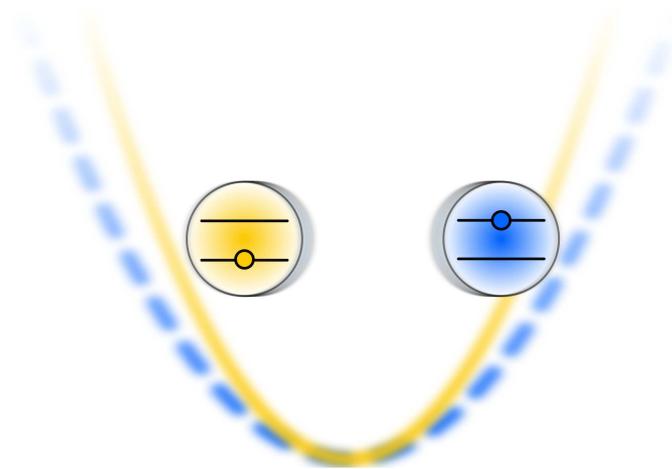}
   \caption{\label{fig:_ionsintrap}%
   As the oscillation frequency of a trapped ion depends on its mass, $\Omega\sim (m)^{-1/2}$, an ion in an excited internal state oscillates at a different frequency, $\Omega^\ast \sim (m+\hbar \omega_0/c^2)^{-1/2}$.}
\end{figure}

Let us consider a single ion, prepared in its ground state and trapped in a harmonic trap with trap angular frequency
$\Omega = \sqrt{k/m}$, where $k$ is the trap stiffness and $m$ denotes the mass of the ion.  Using suitable optical
pulses we can transfer the ion into a superposition of the initial ground state and a suitably chosen metastable 
excited state, separated from the ground state by the energy $\hbar\omega_0$.  If this is performed using a two-photon
transition, then this can be done without affecting the motional state of the ion, which will continue to undergo harmonic
oscillations in the trap.  The ground and excited states, having different masses, will oscillate at different frequencies in
the trap, as depicted in Fig. \ref{fig:_ionsintrap}.  The frequency of oscillations for the excited state, $\Omega^\ast$, and for the ground state, $\Omega$, are 
related by
\begin{equation}
   \Omega^\ast = \sqrt{\frac{k}{m + \hbar \omega_0/c^2}} \approx \Omega \left( 1 - \frac{\hbar \omega_0}{2 m c^2}\right) .
\end{equation}
It follows that even if the motional state is unchanged in the creation of the superposition, the subsequent motion
will become entangled with the internal ionic state, as if the two states were associated with different potentials.
The time taken for the motions of the ground and excited states to first become maximally separated is
\begin{equation}
   T = \frac{\pi}{\Omega - \Omega^\ast} \approx \frac{2\pi}{\Omega} \frac{m c^2}{\hbar \omega_0} ,
\end{equation}
which corresponds to $m c^2/(\hbar \omega_0)$ periods of the ion's motion in the trap.  

To observe this separation of the ground and excited states would certainly be technically demanding but is not a hopeless 
task.  The first requirement is for a long-lived excited state so that the experiment can be performed before
the metastable state decays.  Ions studied for possible frequency standards have been shown to have very long-lived
metastable states.  These include the $^{171}\text{Yb}^+$ ion; its $^2\mathrm{F}_{7/2}$-state is remarkably stable 
with a lifetime of several years \cite{roberts1997observation,roberts2000observation}.  The transition frequency between
this state and the ground state is $\omega_0/(2 \pi) = 642\,\mathrm{THz}$ and this means that $\hbar \omega_0/(m c^2) 
\approx 1\cdot36\times 10^{-11}$.  Hence the maximum spatial separation between the ground and excited states will occur
after about $7\cdot4\times 10^{10}$ motional periods.  For a trap frequency of $1\cdot3 \, \mathrm{MHz}$ \cite{roberts2000observation}
this corresponds to a time of $5\cdot 7 \times 10^4 \mathrm{s}$ or $15$ hours, which is substantially less than the 
excited-state lifetime.  One would, of course, have to keep the trap stable and avoid interruptions to the ion, such as collisions with background gas atoms.

Although challenging, there is a subtle and highly unusual feature of the proposed experiment that makes it worthy of further
consideration.  This is the fact that we are exploring a superposition of \emph{different mass} states rather than energy 
states as the spatial separation of the wave-packets for the ground and excited states depends on the difference in rest masses
for these states.  It is interesting to note that there are good reasons arising from Gallilean invariance why such a superposition 
is not possible in non-relativistic quantum theory, although this restriction does not extend to the relativistic domain
\cite{Weinberg}.

\section{Conclusion}

We have seen how a simple application of ideas from non-relativistic physics leads to a paradox, the existence of a
vacuum friction force.  If we take this force to be a damping of the velocity then we run into a problem, the corresponding
necessity of a preferred frame of absolute rest.  The existence of such a state of absolute rest would be in direct conflict
with special relativity but also with Newton's mechanics and, in particular, with his first law of motion.

The quantitive resolution of the paradox is, as we have seen, that the exited atom loses mass in undergoing spontaneous
decay and that the amount of mass lost is precisely $\hbar\omega_0/c^2$.  When we combine this idea with energy 
quantisation we are led directly to $E = mc^2$, a key consequence of special relativity.  The remarkable feature of this,
however, is no explicitly relativistic ideas were used to derive it; we needed to use only non-relativistic quantum theory, the 
first-order Doppler shift and Bradley's 1729 notion of aberration due to motion.  We may ponder the point at which relativity
sneaked into our analysis or simply marvel at the way in which in physics seems to take care of itself and has no regard for
our attempts to classify parts of it as classical or quantum, or as relativistic or non-relativistic.

\section*{Endnote (SMB)}

Danny Segal was a lovely man, a generous and caring teacher, and a talented, enthusiastic and imaginative physicist.  
He enjoyed, perhaps especially, the absurdities that our chosen discipline throws up from time to time and I 
would very much loved to have had the chance to show to him the one presented above, to benefit
from his wisdom and to see him smile.

\section*{Acknowledgement}

We are grateful to Nils Trautmann, Mohamed Babiker, Jim Cresser and Helmut Ritsch with whom we have enjoyed
many interesting discussions on this topic.

\section*{Disclosure statement}

The authors declare no conflict of interest.

\section*{Funding}

This work was supported by a Royal Society Research Professorship (RP150122) and by the Austrian Science Fund FWF
(Grant No. J 3703-N27).

\section*{Publisher's statement}
This is an Accepted Manuscript of an article published by Taylor \& Francis in \emph{Journal of Modern Optics} on 14 September 2017, available online: \href{http://dx.doi.org/10.1080/09500340.2017.1374482}{http://www.tandfonline.com/10.1080/09500340.2017.1374482}

\end{document}